\documentclass[twocolumn,english,notitlepage,pra,aps,showpacs]{revtex4}
\usepackage{amsmath}
\usepackage{amssymb}
\usepackage{graphicx}
\usepackage{esint}
\usepackage{color}

\pacs{03.65.Ta, 03.65.Ca}

\begin{document}

\title{Heisenberg position-momentum uncertainty relation beyond central
potentials}

\author{\L{}ukasz Rudnicki}

\email{rudnicki@cft.edu.pl}

\affiliation{Center for Theoretical Physics, Polish Academy of Sciences, Aleja
Lotnik{\'o}w 32/46, PL-02-668 Warsaw, Poland}
\begin{abstract}
Recently it was shown in {[}New J. Phys. \textbf{8}, 330 (2006){]}
that the three dimensional Heisenberg uncertainty relation $\sigma_{r}\sigma_{p}\geq3\hbar/2$
might be significantly sharpened if the relevant quantum state describes
the particle in a central potential. I extend that result to the case
of states which are not the eigenstates of the square of the angular
momentum operator. I derive a new lower bound for $\sigma_{r}\sigma_{p}$
which involves the mean value and the variance of the square of the angular momentum operator. 
\end{abstract}
\maketitle
\section{Introduction}

The famous Heisenberg uncertainty relation \cite{heis,kennard} is
one of the most fundamental results of the whole quantum mechanics.
Furthermore, testing various uncertainty relations provides a way
to characterize many important quantum features. For a generic three
dimensional system described in terms of a wave function $\psi\left(\boldsymbol{r}\right)$
the Heisenberg uncertainty relation reads
\begin{equation}
\sigma_{r}\sigma_{p}\geq\frac{3\hbar}{2},\label{Heisenberg}
\end{equation}
where  $\sigma_{r}=\sqrt{\left\langle r^{2}\right\rangle -\left\langle \boldsymbol{r}\right\rangle \cdot\left\langle \boldsymbol{r}\right\rangle}$ and  $\sigma_{p}=\sqrt{\left\langle p^{2}\right\rangle -\left\langle \boldsymbol{p}\right\rangle \cdot\left\langle \boldsymbol{p}\right\rangle}$ denote the standard deviations  of
position $\boldsymbol{r}$ and momentum $\boldsymbol{p}$ variables respectively. 

In a special case where the wave function is an eigenstate of the
square of the angular momentum operator $\hat{L}^{2}\psi\left(\boldsymbol{r}\right)=\hbar^{2}l\left(l+1\right)\psi\left(\boldsymbol{r}\right)$
the lower bound in (\ref{Heisenberg}) was significantly sharpened
in \cite{PabloJesus} (this result has been recently rediscovered
in \cite{Bracher})
\begin{equation}
\sigma_{r}\sigma_{p}=\sqrt{\left\langle r^{2}\right\rangle\left\langle p^{2}\right\rangle}\geq\hbar\left(l+\frac{3}{2}\right).\label{PJ}
\end{equation}
The fact that $\left\langle \boldsymbol{r}\right\rangle =0=\left\langle \boldsymbol{p}\right\rangle$ for all eigenstates of  $\hat{L}^{2}$  was taken into account. We shall also notice that all results presented in \cite{PabloJesus} were derived in a general case of $D$-dimensional systems.

The eigenstates of the $\hat{L}^{2}$ operator play a special role
in atomic physics, where a quantum system is usually assumed to evolve
in a central potential depending only on the distance from the origin
$r^{2}=\boldsymbol{r}\cdot\boldsymbol{r}$. For several central potentials
the product $\sigma_{r}\sigma_{p}$ was analytically derived and checked
numerically in \cite{examples}. Various uncertainty relations \cite{unc1,unc2,unc3,unc4,unc5,unc6,unc7,unc8}
are of special importance in numerical computations related to electronic
structures, in particular to that based on the density functional
theory \cite{functional theory1}, because they provide sensitive
tools to verify physical adequateness of obtained electron densities
\cite{functional theory 2}. However, since the results of numerical
calculations and possible experiments cannot guarantee us that $\psi\left(\boldsymbol{r}\right)$
has all expected properties, we shall ask about the validity of the
uncertainty relation (\ref{PJ}) when $\psi\left(\boldsymbol{r}\right)$
is not exactly the eigenstate of $\hat{L}^{2}$. The aim of this paper
is to give an answer to that question.

Obviously, the standard deviations  $\sigma_{r}$ and  $\sigma_{p}$ are invariant under translations in both position and momentum (transformations $\boldsymbol{r}\mapsto\boldsymbol{r}+\boldsymbol{r}_{0}$ and $\boldsymbol{p}\mapsto\boldsymbol{p}+\boldsymbol{p}_{0}$), however the average value of $\hat{L}^{2}$ is not. Therefore, let me introduce an invariant version of the angular momentum operator ($\boldsymbol{r}$ and $\boldsymbol{p}$ are operators)
\begin{equation}
\hat{\boldsymbol{L}}_{\textrm{inv}}=\left(\boldsymbol{r}-\left\langle \boldsymbol{r}\right\rangle \right)\times\left(\boldsymbol{p}-\left\langle \boldsymbol{p}\right\rangle \right).
\end{equation}
This is the angular momentum in the position and momentum reference frames centered at $\left\langle \boldsymbol{r}\right\rangle$ and $\left\langle \boldsymbol{p}\right\rangle$ respectively. Let me then assume that we have at our disposal the average value
of the square of this angular momentum operator described by
a dimensionless parameter $\mathrm{L}_{\textrm{inv}}$
\begin{equation}
\hbar^{2}\mathrm{L}_{\textrm{inv}}^{2}=\left\langle \hat{L}_{\textrm{inv}}^{2}\right\rangle =\int d^{3}r\,\psi^{*}\left(\boldsymbol{r}\right)\hat{L}_{\textrm{inv}}^{2}\psi\left(\boldsymbol{r}\right).
\label{l}
\end{equation}
In Section \ref{sec:example}, I present a simple example which shows
that when one does not know the state (one knows only the parameter
$\mathrm{L}_{\textrm{inv}}$) one cannot refine the general Heisenberg bound (\ref{Heisenberg}).
This happens because even if $\mathrm{L}^2_{\textrm{inv}}\approx l\left(l+1\right)$
for some $l\in\mathbb{N}$, one can still construct states \textit{laying
far away} from the eigenstate of $\hat{L}_{\textrm{inv}}^{2}$ labeled by the quantum
number $l$. To overcome this issue I shall employ the variance of
$\hat{L}_{\textrm{inv}}^{2}$ and, in addition to $\mathrm{L}_{\textrm{inv}}$, use the dimensionless
parameter 

\begin{equation}
\mathcal{R}_{\textrm{inv}}=\hbar^{-4}\left\langle \left(\hat{L}_{\textrm{inv}}^{2}-\left\langle\hat{L}_{\textrm{inv}}^{2}\right\rangle \right)^{2}\right\rangle .\label{R}
\end{equation}
Of course  $\mathcal{R}_{\textrm{inv}}=0$ only for the eigenstates of $\hat{L}_{\textrm{inv}}^{2}$.
Thus, relatively small values of  $\mathcal{R}_{\textrm{inv}}$ shall justify the
approximation that $\psi\left(\boldsymbol{r}\right)$ is some eigenstate
of $\hat{L}_{\textrm{inv}}^{2}$. 

The main result of this paper is the sharpened version of the Heisenberg
uncertainty relation (\ref{Heisenberg}) 
\begin{equation}
\sigma_{r}\sigma_{p}\geq\frac{3\hbar}{2}+\frac{\hbar\mathrm{L}^{3}_{\textrm{inv}}}{2}\,\frac{\sqrt{\mathrm{L}^{2}_{\textrm{inv}}+4\mathrm{L}^{4}_{\textrm{inv}}+4\mathcal{R}_{\textrm{inv}}}-\mathrm{L}_{\textrm{inv}}}{\mathcal{R}_{\textrm{inv}}+\mathrm{L}^{4}_{\textrm{inv}}},\label{main}
\end{equation}
derived in Section \ref{sec:Derivation}. In the discussion part (Section \ref{sec:discussion}) I show that this new
uncertainty relation links in a continuous manner the previous result
(\ref{PJ}) recovered from (\ref{main}) for $\mathrm{L}_{\textrm{inv}}=\sqrt{l\left(l+1\right)}$
and $\mathcal{R}_{\textrm{inv}}=0$, with the Heisenberg uncertainty relation (\ref{Heisenberg})
achieved when $\mathrm{L}_{\textrm{inv}}=0$ or $\mathcal{R}_{\textrm{inv}}\rightarrow\infty$. We shall notice that for $\left\langle \boldsymbol{r}\right\rangle =0=\left\langle \boldsymbol{p}\right\rangle$ we obtain $\hat{L}_{\textrm{inv}}^{2}=\hat{L}^{2}$ what provides a full correspondence with the case (\ref{PJ}).

\section{The lower bound (\ref{PJ}) beyond eigenstates of $\hat{L}_{\textrm{inv}}^2$}\label{sec:example}

In this Section I show that the constraint on the average value $\mathrm{L}^{2}_\textrm{inv}$
(from now on we put $\hbar=1$) of the $\hat{L}_{\textrm{inv}}^{2}$ operator (or simply $\hat{L}^2$ operator in the reference frame where $\left\langle \boldsymbol{r}\right\rangle =0=\left\langle \boldsymbol{p}\right\rangle$) does
not lead by itself to an inequality stronger than the Heisenberg uncertainty
relation (\ref{Heisenberg}). To this end let me use the function

\begin{equation}
f{}_{l}\left(r,\theta,\varphi\right)=\frac{2^{1+l}}{\pi^{1/4}}\sqrt{\frac{l!}{\left(2l+1\right)!}}r^{l}e^{-r^{2}/2}Y_{l}^{0}\left(\theta,\varphi\right),
\end{equation}
which is the normalized eigenstate of some isotropic harmonic oscillator
with the angular momentum quantum number $l$, the magnetic number
$m=0$, and the principal quantum number $n=0$. Now we shall take
the superposition of the ground state $f_{0}$ (angular momentum number
equal to $0$) with the state $f_{l_{0}}$, for $l_{0}\left(l_{0}+1\right)>\mathrm{L}^{2}_{\textrm{inv}}$  and $l_0>1$
\begin{equation}
\Psi_{l_{0}}=\sqrt{\frac{l_{0}\left(l_{0}+1\right)-\mathrm{L}^{2}_{\textrm{inv}}}{l_{0}\left(l_{0}+1\right)}}f_{0}+\frac{\mathrm{L}_{\textrm{inv}}}{\sqrt{l_{0}\left(l_{0}+1\right)}}f_{l_{0}}.\label{example}
\end{equation}
 The state (\ref{example}) satisfies $\left\langle \boldsymbol{r}\right\rangle =0$, $\left\langle \boldsymbol{p}\right\rangle =0$ and $\left\langle \hat{L}_{\textrm{inv}}^{2}\right\rangle=\left\langle \hat{L}^{2}\right\rangle=\mathrm{L}^{2}_{\textrm{inv}}$, but 
\begin{equation}
\sigma_{r}\sigma_{p}=\sqrt{\left\langle r^{2}\right\rangle\left\langle p^{2}\right\rangle}=\frac{3}{2}+\frac{\mathrm{L}^{2}_{\textrm{inv}}}{l_{0}+1}\;\underset{l_{0}\rightarrow\infty}{\longrightarrow}\;\frac{3}{2}.\label{varex}
\end{equation}
This observation means that keeping $\left\langle \hat{L}_{\textrm{inv}}^{2}\right\rangle $
constant we are able to go arbitrarily close to the ground state $f_{0}$
taking arbitrarily large $l_{0}$. However, for the example state
(\ref{example}) we can check that 
\begin{equation}
\mathcal{R}_{\textrm{inv}}=\mathrm{L}^{2}_{\textrm{inv}}\left(l_{0}\left(l_{0}+1\right)-\mathrm{L}^{2}_{\textrm{inv}}\right)\underset{l_{0}\rightarrow\infty}{\longrightarrow}\infty,
\end{equation}
 which explains, why the Heisenberg lower bound $3/2$ can be asymptotically
reached in (\ref{varex}). 

\section{Proof of the uncertainty relation (\ref{main}) based on calculus of variations}\label{sec:Derivation}

A main ingredient of my derivation shall be the variational approach
used recently in \cite{ibbZ} to prove a new Heisenberg-like uncertainty
relation for photons. The most important advantage of this method
is that one does not need to rely on commutation relations between
the conjugate variables. In the first step I
define the following functionals:
\begin{equation}
X^{2}\left[\psi^{*},\psi\right]=\int d^{3}r\, r^{2}\psi^{*}\left(\boldsymbol{r}\right)\psi\left(\boldsymbol{r}\right),
\end{equation}
\begin{equation}
P^{2}\left[\psi^{*},\psi\right]=\int d^{3}r\,\psi^{*}\left(\boldsymbol{r}\right)\left(-\triangle\right)\psi\left(\boldsymbol{r}\right),
\end{equation}
where $\triangle$ is a three-dimensional Laplacian and we put $\hbar=1$.

\subsection{Ordinary Heisenberg uncertainty relation}\label{subsec:ordinary}

Following the idea presented in \cite{ibbZ} I shall briefly describe
the variational method using the example of the Heisenberg uncertainty
relation (\ref{Heisenberg}). However, I modify this derivation and
use the Lagrange multiplier to include the normalization condition.
To prove (\ref{Heisenberg}) one can start with solving the following variational
equation
\begin{equation}
\!\frac{\delta}{\delta\psi^{*}}\!\left[X^{2}\left[\psi^{*}\!,\psi\right]P^{2}\left[\psi^{*}\!,\psi\right]-\!\lambda\!\left(\int d^{3}r\,\psi^{*}\!\!\left(\boldsymbol{r}\right)\psi\left(\boldsymbol{r}\right)-\!\!1\right)\!\right]=0,\label{variational1}
\end{equation}
where $\lambda$ is the Lagrange multiplier related to the normalization
constraint of the wave function. The equation (\ref{variational1})
gives
\begin{equation}
\left(\frac{\omega^{2}}{X^{2}}r^{2}-X^{2}\triangle-\lambda\right)\psi\left(\boldsymbol{r}\right)=0,
\label{hadd}
\end{equation}
where we denote $\omega=XP$. In the next step we impose the normalization condition (we multiply (\ref{hadd}) by $\psi^*$  and integrate over $d^{3}r$) and find that  $\lambda=2\omega^{2}$.
Finally, we introduce a dimensionless variable $\boldsymbol{\xi}=\boldsymbol{r}/X$
and obtain the Schr\"odinger equation for the three dimensional isotropic
harmonic oscillator 
\begin{equation}
\left(-\frac{1}{2}\triangle+\frac{\omega^{2}\xi^{2}}{2}\right)\psi\left(\boldsymbol{\xi}\right)=\omega^{2}\psi\left(\boldsymbol{\xi}\right).\label{Schrodinger1}
\end{equation}
The eigenvalues of the \textit{left hand side} of (\ref{Schrodinger1})
are $\omega\left(n+3/2\right)$, $n\in\mathbb{N}$, thus, the eigenvalue
equation for (\ref{Schrodinger1}) gives 
\begin{equation}
\omega\left(n+\frac{3}{2}\right)=\omega^{2}\quad\Longrightarrow\quad\omega=n+\frac{3}{2}.\label{eigenvalue1}
\end{equation}
The smallest possible value $\omega_{min}$ of $\omega$ in (\ref{eigenvalue1})
is equal to $3/2$ for $n=0$ and when $\psi\left(\boldsymbol{\xi}\right)$
is the ground state. This result provides us the inequality
\begin{equation}
XP\geq\omega_{min}=\frac{3}{2}. \label{part}
\end{equation}

Now we choose the coordinate and momentum reference frames such that  $\left\langle \boldsymbol{r}\right\rangle =0=\left\langle \boldsymbol{p}\right\rangle$. In this specific frame we have $X=\sigma_r$ and $P=\sigma_p$, and our result (\ref{part}) reads
\begin{equation}
\sigma_{r}\sigma_{p}\geq\frac{3}{2}. \label{part2}
\end{equation}
The observation that (\ref{part2}) is invariant under translations (is valid in all reference frames) completes the proof of (\ref{Heisenberg}). In other words, in order to prove the Heisenberg uncertainty relation (\ref{Heisenberg}) one needs to show that $\sqrt{\left\langle r^{2}\right\rangle\left\langle p^{2}\right\rangle}\geq 3/2$, when $\left\langle \boldsymbol{r}\right\rangle =0=\left\langle \boldsymbol{p}\right\rangle$.

\subsection{Derivation including constraints related to angular momentum}

In order to prove the main result of this paper (\ref{main}) I shall
use the method described in \ref{subsec:ordinary} together with two
additional constraints: $\left\langle \hat{L}^{2}\right\rangle\equiv\mathrm{L}^2=\textrm{const}$ and $\left\langle \left(\hat{L}^{2}-\left\langle\hat{L}^{2}\right\rangle \right)^{2}\right\rangle\equiv\mathcal{R}=\textrm{const}$. Note that in this part of the derivation I use the ordinary angular momentum operator $\hat{\boldsymbol{L}}$.
The variational equation in that case reads ($F=\mathcal{R}+\mathrm{L}^{4}$) 
\begin{widetext}
\begin{equation}
\frac{\delta}{\delta\psi^{*}}\!\left[{X^{2}}\left[\psi^{*}\!,\psi\right]{P^{2}}\left[\psi^{*}\!,\psi\right]-\!\lambda\!\left(\int\!\! d^{3}r\,\psi^{*}\!\!\left(\boldsymbol{r}\right)\psi\left(\boldsymbol{r}\right)-\!\!1\right)\!+2\eta\!\left(\int\!\! d^{3}r\,\psi^{*}\!\!\left(\boldsymbol{r}\right)\!\hat{L}^{2}\psi\!\left(\boldsymbol{r}\right)\!-\!\mathrm{L}^{2}\!\right)\!+\!2\Lambda\!\left(\int\!\! d^{3}r\,\psi^{*}\!\!\left(\boldsymbol{r}\right)\!\hat{L}^{4}\psi\!\left(\boldsymbol{r}\right)\!-\! F\right)\!\right]=0,\label{variational2}
\end{equation}
\end{widetext} and leads to the equation
\begin{equation}
\left(\frac{\omega^{2}}{X^{2}}r^{2}-X^{2}\triangle-\lambda+2\eta\hat{L}^{2}+2\Lambda\hat{L}^{4}\right)\psi\left(\boldsymbol{r}\right)=0.
\end{equation}
The parameters $\lambda$, $\eta$ and $\Lambda$ in (\ref{variational2})
play the role of Lagrange multipliers and the factor $-2$ before
$\eta$ and $\Lambda$ was introduced for further convenience. Imposing
the normalization constraint we find that $\lambda=2\omega^{2}+2\eta\mathrm{L}^{2}+2\Lambda F$.
Thus, we have to solve the following counterpart of the Schr\"odinger
equation (\ref{Schrodinger1}) 
\begin{equation}
\left(-\frac{1}{2}\triangle+\frac{\omega^{2}\xi^{2}}{2}+\eta\hat{L}^{2}+\Lambda\hat{L}^{4}\right)\psi\left(\boldsymbol{\xi}\right)=\mathcal{E}\psi\left(\boldsymbol{\xi}\right),\label{hamilton}
\end{equation}
where $\mathcal{E}=\omega^{2}+\eta\mathrm{L}^{2}+\Lambda F$. 

\subsubsection{Solutions to the eigenproblem}

In the first step we notice that all eigenstates $\Phi\left(\boldsymbol{\xi}\right)$
of the isotropic harmonic oscillator \cite{Quanta}
\begin{equation}
\left(-\frac{1}{2}\triangle+\frac{\omega^{2}\xi^{2}}{2}\right)\Phi\left(\boldsymbol{\xi}\right)=\omega\left(\frac{3}{2}+2n+l\right)\Phi\left(\boldsymbol{\xi}\right).
\end{equation}
are also solutions of (\ref{hamilton}). However, taking $\psi\left(\boldsymbol{\xi}\right)$
to be one of these eigenstates, we will not be able to fulfill the
constraints on angular momentum (in particular all eigenstates $\Phi\left(\boldsymbol{\xi}\right)$
have $\mathcal{R}=0$). Thus, in order to solve the equation (\ref{hamilton})
together with the constraints we shall look for the solution of (\ref{hamilton})
in the form of a superposition of the eigenstates of the harmonic
oscillator
\begin{equation}
\psi\left(\boldsymbol{\xi}\right)=\sum_{i=1}^{N}C_{i}\Phi_{i}\left(\boldsymbol{\xi}\right),\qquad\sum_{i=1}^{N}\left|C_{i}\right|^{2}=1.\label{sum}
\end{equation}
We assume that each two states $\Phi_{i}$ and $\Phi_{j}$ (for $i\neq j$) in (\ref{sum}) differ by at least one quantum number, i.e. $n_i\neq n_j$ or $l_i \neq l_j$. Thus, all states $\Phi_{i}\left(\boldsymbol{\xi}\right)$
in the superposition (\ref{sum}) are independently the eigenstates
of (\ref{hamilton}) with the energies 
\begin{equation}
\mathcal{E}_{i}=\omega\left(\frac{3}{2}+2n_{i}+l_{i}\right)+\eta l_{i}\left(l_{i}+1\right)+\Lambda l_{i}^{2}\left(l_{i}+1\right)^{2}.\label{eigen}
\end{equation}
In that way we obtain a set of $N$ independent equations that must
be simultaneously satisfied
\begin{equation}
\mathcal{E}_{i}=\mathcal{E},\qquad i=1,2,\ldots,N.\label{problem}
\end{equation}
But, in our problem we have only three constants to specify: $\mathcal{E}$,
$\eta$, $\Lambda$ (it is important to notice that amplitudes $C_{i}$
do not appear in (\ref{problem})), thus, allowed solutions (\ref{sum})
of (\ref{hamilton}) can be the superposition of at most three eigenstates
of the harmonic oscillator. In other words only the value $N=3$ allows
us to find the solutions $\eta\left(\omega,n_{i},l_{i}\right)$, $\Lambda\left(\omega,n_{i},l_{i}\right)$
and $\mathcal{E}\left(\omega,n_{i},l_{i}\right)$. I will not write
them down explicitly, but I restrict myself to give in \ref{subsec:final}
a simple derivation of $\omega$ consistent with these solutions.

It might appear unusual that we have not specified the Lagrange multipliers
using the related constraints, but during the process of solving the
variational equation (\ref{variational2}). However, we still have
at our disposal three amplitudes $C_{1}$, $C_{2}$ and $C_{3}$ and
we shall use these coefficients to fulfill the remaining constraints. 

\subsubsection{Solutions to the constraints}

For the sake of simplicity let me introduce the following notation:
$\alpha=l_{1}\left(l_{1}+1\right)$, $\beta=l_{2}\left(l_{2}+1\right)$,
$\gamma=l_{3}\left(l_{3}+1\right)$. The three constraints we have
imposed lead to three equations for the probabilities $\left|C_{i}\right|^{2}$:\begin{subequations}
\begin{equation}
\left|C_{1}\right|^{2}+\left|C_{2}\right|^{2}+\left|C_{3}\right|^{2}=1,\label{eq1}
\end{equation}
\begin{equation}
\alpha\left|C_{1}\right|^{2}+\beta\left|C_{2}\right|^{2}+\gamma\left|C_{3}\right|^{2}=\mathrm{L}^{2},
\end{equation}
\begin{equation}
\alpha^{2}\left|C_{1}\right|^{2}+\beta^{2}\left|C_{2}\right|^{2}+\gamma^{2}\left|C_{3}\right|^{2}=F.\label{eq3}
\end{equation}
\end{subequations}The solutions of (\ref{eq1}-\ref{eq3}) are: \begin{subequations}
\begin{equation}
\left|C_{1}\right|^{2}=\frac{\beta\gamma-\mathrm{L}^{2}\left(\beta+\gamma\right)+F}{\left(\alpha-\beta\right)\left(\alpha-\gamma\right)},\label{sol1}
\end{equation}
\begin{equation}
\left|C_{2}\right|^{2}=\frac{\alpha\gamma-\mathrm{L}^{2}\left(\alpha+\gamma\right)+F}{\left(\beta-\alpha\right)\left(\beta-\gamma\right)},
\end{equation}
\begin{equation}
\left|C_{3}\right|^{2}=\frac{\alpha\beta-\mathrm{L}^{2}\left(\alpha+\beta\right)+F}{\left(\gamma-\alpha\right)\left(\gamma-\beta\right)}.\label{sol3}
\end{equation}
\end{subequations}

In that way we have specified the moduli of $C_{1}$, $C_{2}$ and
$C_{3}$ coefficients, but their phases might be arbitrary. The solutions
(\ref{sol1}-\ref{sol3}) possess a permutational symmetry (a permutation
of $l_{i}$ produces the permutation of $\left|C_{i}\right|$), however
at this point we shall decide about some hierarchy between the $l_{i}$
numbers. Let me choose $l_{1}\geq l_{2}\geq l_{3}$ what means that
$\alpha\geq\beta\geq\gamma$. Since the solutions (\ref{sol1}-\ref{sol3})
must be positive we obtain the following conditions for $\alpha$,
$\beta$ and $\gamma$:\begin{subequations}
\begin{equation}
\beta\gamma-\mathrm{L}^{2}\left(\beta+\gamma\right)+F\geq0,\label{in1}
\end{equation}
\begin{equation}
\alpha\gamma-\mathrm{L}^{2}\left(\alpha+\gamma\right)+F\leq0,
\end{equation}
\begin{equation}
\alpha\beta-\mathrm{L}^{2}\left(\alpha+\beta\right)+F\geq0.\label{in3}
\end{equation}
\end{subequations}The conditions (\ref{in1}-\ref{in3}) might be
also presented in terms of two, mutually exclusive cases:
\begin{equation}
0\leq\gamma\leq\mathrm{L}^{2}\leq\beta\leq\frac{F-\gamma\mathrm{L}^{2}}{\mathrm{L}^{2}-\gamma}\leq\alpha,\label{dom1}
\end{equation}
or
\begin{equation}
0\leq\gamma\leq\beta\leq\mathrm{L}^{2}\leq\frac{F-\gamma\mathrm{L}^{2}}{\mathrm{L}^{2}-\gamma}\leq\alpha\leq\frac{F-\beta\mathrm{L}^{2}}{\mathrm{L}^{2}-\beta}.\label{dom2}
\end{equation}
In fact, from the beginning we could expect that either one or two
parameters among $\left(\alpha,\beta,\gamma\right)$ shall be greater
than $\mathrm{L}^{2}$. 

\subsection{Final uncertainty relation}\label{subsec:final}

Since all constraints are fulfilled we are able to derive in a simple
way the coefficient $\omega$ consistent with the equations (\ref{problem}).
To this end I write down these equations explicitly (for $i\in\left\{ 1,2,3\right\} $)
\begin{equation}
\omega\!\left(\frac{3}{2}+2n_{i}\!+l_{i}\!\right)+\eta l_{i}\!\left(l_{i}+1\right)+\Lambda l_{i}^{2}\!\left(l_{i}+1\right)^{2}=\!\omega^{2}+\eta\mathrm{L}^{2}+\!\Lambda F,\label{eq36}
\end{equation}
multiply $i$-th equation by $\left|C_{i}\right|^{2}/\omega$ and
sum up over $i$. Due to the conditions (\ref{eq1}-\ref{eq3}) all
the terms with the Lagrange multipliers $\eta$ and $\Lambda$ present
on both sides of (\ref{eq36}) cancel, and we immediately obtain the
solution 
\begin{equation}
\omega\left(n_{i},l_{i}\right)=\frac{3}{2}+\sum_{i=1}^{3}\left|C_{i}\right|^{2}\left(2n_{i}+l_{i}\right).
\end{equation}

The final step is to find values of the quantum numbers $n_{i}$ and
$l_{i}$ that minimize $\omega$. Since the probabilities $\left|C_{i}\right|^{2}$
do not depend on $n_{i}$ we shall take the lowest levels $n_{i}=0$,
as in the case of the ordinary Heisenberg uncertainty relation. An
answer to the question which values of the parameters $l_{i}$ minimize
the function $\omega\left(l_{i}\right)\equiv\omega\left(0,l_{i}\right)$
is more subtle, because we have assumed that $l_{i}\in\mathbb{N}$.
However, for a moment we shall treat $l_{i}$ as continuous parameters
(we assume that $l_{i}\in\mathbb{R}$) and calculate the following
derivatives:
\begin{subequations}
\begin{equation}
\frac{\partial\omega\left(l_{i}\right)}{\partial l_{1}}=\frac{\left(l_{1}-l_{2}\right)\left(l_{1}-l_{3}\right)\left(S+l_{1}\right)}{Q_{12}Q_{13}Q_{23}}\left|C_{1}\right|^{2}\geq0,
\end{equation}
\begin{equation}
\frac{\partial\omega\left(l_{i}\right)}{\partial l_{2}}=\frac{\left(l_{2}-l_{1}\right)\left(l_{2}-l_{3}\right)\left(S+l_{2}\right)}{Q_{12}Q_{13}Q_{23}}\left|C_{2}\right|^{2}\leq0,
\end{equation}
\begin{equation}
\frac{\partial\omega\left(l_{i}\right)}{\partial l_{3}}=\frac{\left(l_{3}-l_{1}\right)\left(l_{3}-l_{2}\right)\left(S+l_{3}\right)}{Q_{12}Q_{13}Q_{23}}\left|C_{3}\right|^{2}\geq0,
\end{equation}
\end{subequations}
where $S=2+l_{1}+l_{2}+l_{3}$ and $Q_{jk}=1+l_{j}+l_{k}$. 

We have defined $\omega_{min}$ as a minimal value of $\omega\left(l_{i}\right)$
with the assumption that $l_{i}\in\mathbb{N}$. Thus, since $\omega\left(l_{i}\right)$
increases with $l_{1}$ and $l_{3}$ and decreases with $l_{2}$ we
can obtain a lower bound $\Omega\leq\omega_{min}$ taking the smallest
possible values of $l_{1}$ and $l_{3}$ and the largest value of
$l_{2}$, according to the ranges (\ref{dom1}, \ref{dom2}). This
means that the optimal values of $l_{i}\in\mathbb{R}$ are: 
\begin{equation}
l_{1}=l_{2}=\frac{\sqrt{4F+\mathrm{L}^{2}}}{2\mathrm{L}}-\frac{1}{2},\qquad l_{3}=0,
\end{equation}
and lead to the result
\begin{equation}
\Omega\left(\mathrm{L},\mathcal{R}\right)=\frac{3}{2}+\frac{\mathrm{L}^{4}}{2F\left(\mathrm{L},\mathcal{R}\right)}\left(\sqrt{1+\frac{4F\left(\mathrm{L},\mathcal{R}\right)}{\mathrm{L}^{2}}}-1\right).\label{main2}
\end{equation}
When we substitute in (\ref{main2}) the value $F\left(\mathrm{L},\mathcal{R}\right)=\mathcal{R}+\mathrm{L}^{4}$
we obtain the inequality
\begin{equation}
XP\geq\omega_{min}\geq\Omega=\frac{3}{2}+\frac{\mathrm{L}^{3}}{2}\,\frac{\sqrt{\mathrm{L}^{2}+4\mathrm{L}^{4}+4\mathcal{R}}-\mathrm{L}}{\mathcal{R}+\mathrm{L}^{4}}.\label{pomocnik0}
\end{equation}
Similarly to the result (\ref{part}), the uncertainty relation (\ref{pomocnik0}) in the reference frame defined by $\left\langle \boldsymbol{r}\right\rangle =0=\left\langle \boldsymbol{p}\right\rangle$ reads
\begin{equation}
\sigma_{r}\sigma_{p}\geq\frac{3}{2}+\frac{\mathrm{L}^{3}}{2}\,\frac{\sqrt{\mathrm{L}^{2}+4\mathrm{L}^{4}+4\mathcal{R}}-\mathrm{L}}{\mathcal{R}+\mathrm{L}^{4}},\label{pomocnik}
\end{equation}
where now, since $\hat{\boldsymbol{L}}=\hat{\boldsymbol{L}}_{\textrm{inv}}$, we can replace $\hat{\boldsymbol{L}}$ by $\hat{\boldsymbol{L}}_{\textrm{inv}}$ and the couple of parameters $(\mathrm{L},\mathcal{R})$  by $(\mathrm{L}_\textrm{inv}, \mathcal{R}_\textrm{inv})$.
Finally we shall use the fact that $\hat{\boldsymbol{L}}_{\textrm{inv}}$ is invariant under translations both in positions and momenta, thus the uncertainty relation (\ref{pomocnik}) is also invariant and the main result (\ref{main}) of this paper is proven.

In fact, we are able to improve immediately the inequality (\ref{main})
because we can take an exact, but more complicated value $\omega_{min}$
instead of its lower bound $\Omega$. If $l_{i}\in\mathbb{N}$, then
the optimal values of $l_{i}$ are: $l_{3}=0$, and 
\begin{equation}
l_{1}=\left\lceil \frac{\sqrt{4F+\mathrm{L}^{2}}}{2\mathrm{L}}-\frac{1}{2}\right\rceil ,\qquad l_{2}=\left\lfloor \frac{\sqrt{4F+\mathrm{L}^{2}}}{2\mathrm{L}}-\frac{1}{2}\right\rfloor ,
\end{equation}
where $\left\lceil \cdot\right\rceil $ and $\left\lfloor \cdot\right\rfloor $
denote the integer valued ceiling and floor functions respectively
\footnote{These functions may be defined as: $\left\lceil x\right\rceil =\min\left(m\in\mathbb{Z}:\; x\leq m\right)$
and $\left\lfloor x\right\rfloor =\max\left(n\in\mathbb{Z}:\; x\geq n\right)$.%
}. Finally we have
\begin{equation}
\sigma_{r}\sigma_{p}\geq\omega_{min}=W\left(\frac{\sqrt{4\mathcal{R}_{\textrm{inv}}+4\mathrm{L}^{4}_{\textrm{inv}}+\mathrm{L}^{2}_{\textrm{inv}}}}{2\mathrm{L}_{\textrm{inv}}}-\frac{1}{2}\right),
\end{equation}
where the function $W\left(x\right)$ reads ($F_{\textrm{inv}}=\mathcal{R}_{\textrm{inv}}+\mathrm{L}^{4}_{\textrm{inv}}$)
\begin{equation}
W\!\left(x\right)=\!\frac{\mathrm{L}^{2}_{\textrm{inv}}\left(\!\left(1+\left\lceil x\right\rceil \right)^{2}+\left(1+\left\lfloor x\right\rfloor \right)^{2}+\left\lceil x\right\rceil \!\left\lfloor x\right\rfloor -1\!\right)-\!F_{\textrm{inv}}}{\left(1+\left\lceil x\right\rceil \right)\left(1+\left\lfloor x\right\rfloor \right)}.
\end{equation}
One can check that this result is actually a minor improvement of (\ref{main}), moreover the function $W(x)$ is not even continuous.

\begin{figure}
\includegraphics[scale=1]{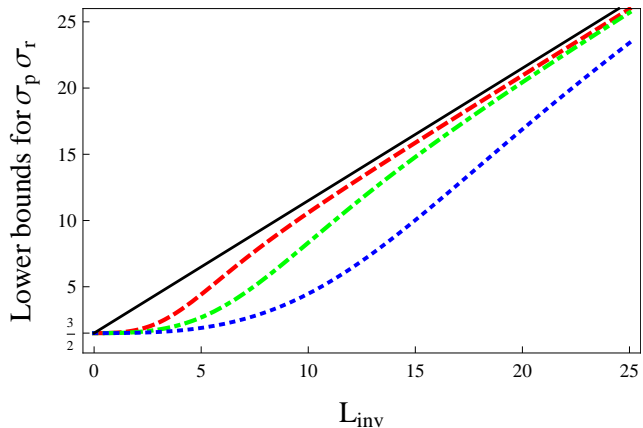}
\caption{(Color online) The dependence on $\mathrm{L}_{\textrm{inv}}$ of the new bound (\ref{main}) for
some fixed values of $\mathcal{R}_{\textrm{inv}}$: $\mathcal{R}_{\textrm{inv}}=10^{3}$ - red/dashed,
$\mathcal{R}_{\textrm{inv}}=10^{4}$ - green/dashed-dotted, $\mathcal{R}_{\textrm{inv}}=10^{5}$
- blue/dotted. The black line represents the reference bound (\ref{PJ}).
All bounds are above the Heisenberg bound $3/2$. \label{fig:The-dependence-on}}
\end{figure}

\section{Discussion}\label{sec:discussion}
In this Section I would like to present the connection of the new
uncertainty relation (\ref{main}) with the previous results (\ref{Heisenberg})
and (\ref{PJ}). First of all we shall note, looking at the expression
(\ref{pomocnik0}), that 
\begin{equation}
\Omega\left(0,\mathcal{R}\right)=\frac{3}{2},\:\textrm{ and }\quad\lim_{\mathcal{R}\rightarrow\infty}\Omega\left(\mathrm{L},\mathcal{R}\right)=\frac{3}{2}.
\end{equation}
These results mean that in both cases, when the angular momentum is
$\mathrm{L}=0$ or when the variance $\mathcal{R}$ is very large,
we have no improvement of the ordinary Heisenberg uncertainty relation
(\ref{Heisenberg}). This conclusion is in a full agreement with logical
expectations. Furthermore, it is easy to check that 
\begin{equation}
\Omega\left(\sqrt{l\left(l+1\right)},0\right)=\frac{3}{2}+l,
\end{equation}
what coincides with the uncertainty relation (\ref{PJ}). The value
$\mathcal{R}=0$ means that the state is a true eigenstate of $\hat{L}^2$ (or $\hat{L}_{\textrm{inv}}^{2}$ in a general reference frame)
and the average value $\mathrm{L}^{2}$ must be equal to $l\left(l+1\right)$
where $l$ is related quantum number. Fig. \ref{fig:The-dependence-on}
summarizes these observations. 

\section{Conclusions}

I this paper I have discussed the Heisenberg uncertainty relation
for position and momentum with additional information about angular
momentum. I derived new lower bound for the product of standard deviations
$\sigma_{r}\sigma_{p}$ which depends on the average value and the
variance of the $\hat{L}_{\textrm{inv}}^{2}$ operator. This operator is a square of the angular momentum operator defined in the reference frame where $\left\langle \boldsymbol{r}\right\rangle =0=\left\langle \boldsymbol{p}\right\rangle$. I showed that the relation
(\ref{main}) links the ordinary Heisenberg uncertainty relation (\ref{Heisenberg})
with stronger relation (\ref{PJ}) valid for the eigenstates of $\hat{L}^{2}_{\textrm{inv}}$. 

\label{sec:conclusions} 
\begin{acknowledgments}
I am especially grateful to Iwo Bia\l{}ynicki-Birula for long discussions and several important
comments. Furthermore, I would like to thank all members of the Atomic Physics Group from University
of Granada for their kind hospitality and fruitful discussions. Finally, the referee deserves thanks for his/her accurate remarks, that help me clarify the main message of this paper.
Financial support by the Grant
of the Polish Ministry of Science and Higher Education for years 2010-2012
is gratefully acknowledged.\end{acknowledgments}

\end{document}